\documentclass[showpacs,preprintnumbers,
amsmath,amssymb,aps,prd]{revtex4}
\usepackage{amsfonts}
\usepackage{amsfonts,graphics,epsfig,subfigure}
\usepackage{float}
\begin{document}

\title{Joule-Thomson expansion of $d$-dimensional charged AdS black holes}
\author{Jie-Xiong Mo\footnote{mojiexiong@gmail.com(corresponding
author)}, Gu-Qiang Li, Shan-Quan Lan, Xiao-Bao Xu}
\affiliation{Institute of Theoretical Physics, Lingnan Normal University, Zhanjiang, 524048, Guangdong, China}

\begin{abstract}
Effects of the dimensionality on the Joule-Thomson expansion are discussed in detail by considering the case of $d$-dimensional charged AdS black holes. Specifically, we investigate three important aspects characteristic of the Joule-Thomson expansion. Namely, the Joule-Thomson coefficient, the inversion curves and the isenthalpic curves. We utilize two different approaches to derive the explicit expression of the Joule-Thomson coefficient and show that both approaches are consistent with each other. The divergent point and the zero point of the Joule-Thomson coefficient are discussed. The former is shown to reveal the information of Hawking temperature while the latter is depicted through the so-called inversion curves. Fine structures of the inversion curves are disclosed in the cases $d>4$. At low pressure, the inversion temperature increases with the dimensionality $d$ while at high pressure it decreases with $d$. The ratio between minimum inversion temperature $T_{min}$ and the critical temperature $T_c$ is discussed with its explicit expression obtained for $d>4$. Surprisingly, it is shown that the ratio is not always equal to $1/2$ but decreases with the dimensionality $d$. Moreover, isenthalpic curves of $d>4$ are shown to expand toward higher pressure when the dimensionality $d$ increases.
\end{abstract}
  \maketitle

\section{Introduction}

    Black hole thermodynamics has long been both an exciting and challenging topic ever since the pioneer works~\cite{Bekenstein,Hawking1,Bardeen}. Thanks to the great contribution of Bekenstein, Hawking and their collaborators, black holes are identified as thermodynamic systems which have temperature and entropy. Unfortunately, Hawking passed away recently, causing inestimable loss to the theoretical physics research.

    Among the researches in the field of black hole thermodynamics, thermodynamics of AdS (Anti-de Sitter) black holes has attracted a lot of attention. The famous Hawking-Page phase transition refers to the transition between the Schwarzschild AdS black hole and the thermal AdS space~\cite{Hawking2}. Close relation between charged AdS black holes and the liquid-gas system was observed by Chamblin et al. and rich phase structures of RN (Reissner Nordstr\"{o}m)-AdS black holes was disclosed \cite{Chamblin1,Chamblin2}. This relation was further enhanced by the discovery \cite{Kubiznak} in the extended phase space. By showing the existence of the reentrant phase transition \cite{Gunasekaran,Altamirano} and triple point \cite{Sherkatghanad}, it was even proposed that AdS black holes behave similarly to the ordinary thermodynamic systems. Huge amount of works has emerged in this field in recent years that we can not list them here. The readers who are interested can refer to the most recent review \cite{Kubiznak2} and references therein.

    Treating the cosmological constant as a variable and identifying it as thermodynamic pressure, a variety of interesting applications have been suggested in literatures. One typical example is the concept of holographic heat engine which was creatively raised by Johnson \cite{Johnson1}. It provides one way to extract mechanical work from black holes. Wei and Liu further generalized this approach to the Rankine cycle and proposed the amazing idea of implementing black holes as efficient power plant \cite{weishaowen}. Another interesting example is the the Joule-Thomson expansion (or Joule-Thomson effect) of black holes, which was investigated in Ref. \cite{Aydiner1} for the first time. Inversion temperature and curves of charged AdS black holes were studied, with similarities and differences between van der Waals fluids and charged AdS black holes compared. This original research was soon generalized to the Kerr-AdS black holes \cite{Aydiner2}, holographic superfluids \cite{Yogendran} and quintessence RN-AdS Black Holes \cite{Ghaffarnejad}.

    In this paper, we would like to generalize the current research of Joule-Thomson expansion to the case of $d$-dimensional charged AdS black holes. The motivation is as follows. Firstly, as stated above, it is commonly recognized by the community that there exists close relation between AdS black holes and ordinary thermodynamic systems. Joule-Thomson expansion, an important procedure in classical thermodynamics, is certainly deserved to be further investigated. It describes the temperature change of a real gas or liquid  when it is forced through a valve or porous plug. This procedure is called a throttling process or Joule-Thomson process. It has widespread applications in thermal machines such as heat pumps, air conditioners and liquefiers. Secondly, choosing the $d$-dimensional charged AdS black holes has its own right. On the one hand, it will help reveal the effect of dimensionality on the Joule-Thomson expansion of black holes. On the other hand, the research in this paper will help further understand the thermodynamics of $d$-dimensional charged AdS black holes. Last but not the least, in both the cases of RN-AdS black holes and Kerr-AdS black holes, it was observed that the ratio between minimum inversion temperature $T_{min}$ and the critical temperature $T_c$ is $1/2$. One may wonder whether this is a universal finding or this observation is only valid in four-dimensional space-time. And we are curious to check this issue in this paper.

    The organization of this paper is as follows. In Sec.\ref{Sec2}, we will present a short review on thermodynamics of $d$-dimensional charged AdS black holes. Joule-Thomson Expansion of $d$-dimensional charged AdS black holes will be investigated in detail in Sec.\ref {Sec3}. At last, we will give a brief summary in Sec.\ref {Sec4}.

\section{A brief review on thermodynamics of $d$-dimensional charged AdS black holes}
\label{Sec2}

The Einstein-Maxwell-anti-de Sitter action for $d$ dimensional space-time reads~\cite{Chamblin1,Chamblin2}
\begin{equation}
I_{EM}=-\frac{1}{16\pi G}\int_M d^dx\sqrt{-g}\left[R-F^2+\frac{(d-1)(d-2)}{l^2}\right],\label{1}
\end{equation}%
where $G$ can be set to one and $l$ denotes the characteristic length scale. The black hole solution of the above action is the famous $d$-dimensional Reissner-Nordstr\"{o}m-anti-de Sitter black hole, whose metric takes the form ~\cite{Gunasekaran}
\begin{equation}
ds^2=-f(r)dt^2+\frac{dr^2}{f(r)}+r^2d\Omega_{d-2}^2,\label{2}
\end{equation}%
where
\begin{equation}
f(r)=1-\frac{m}{r^{d-3}}+\frac{q^2}{r^{2(d-3)}}+\frac{r^2}{l^2}.\label{3}
\end{equation}%
$d\Omega_{d-2}^2$ denotes the metric on the round unit $(d-2)$ sphere while $m$ and $q$ are parameters related to the ADM mass and the electric charge respectively as ~\cite{Chamblin1}
\begin{eqnarray}
M&=&\frac{\omega_{d-2}(d-2)}{16\pi}m,\label{4}
\\
Q&=&\frac{\omega_{d-2}\sqrt{2(d-2)(d-3)}}{8\pi}q.\label{5}
\end{eqnarray}%
$\omega_{d-2}$ denotes the volume of the unit $(d-2)$-sphere and can be obtained from the following relation
\begin{equation}
\omega_d=\frac{2\pi^{\frac{d+1}{2}}}{\Gamma(\frac{d+1}{2})}.\label{6}
\end{equation}%

The thermodynamics of $d$-dimensional charged AdS black holes has been carefully investigated in Ref. \cite{Chamblin1,Chamblin2}, where the Hawking temperature, entropy and electric potential were obtained as
\begin{eqnarray}
T&=&\frac{1}{\beta}=\frac{f'(r_+)}{4\pi}=\frac{d-3}{4\pi r_+}\left(1-\frac{q^2}{r_+^{2(d-3)}}+\frac{d-1}{d-3} \frac{r_+^2}{l^2}\right),\label{7}
\\
S&=&\frac{\omega_{d-2}r_+^{d-2}}{4}, \label{8}
\\
\Phi&=&\sqrt{\frac{d-2}{2(d-3)}}\frac{q}{r_+^{d-3}}, \label{9}
\end{eqnarray}%
where $\beta$ represents the period of the Euclidean section.

In the extended phase space where the cosmological constant is identified as thermodynamic pressure through the definition $P=-\frac{\Lambda}{8\pi}=\frac{(d-1)(d-2)}{16\pi l^2}$, $P-V$ criticality was elegantly studied in Ref.~\cite{Gunasekaran} with the critical physical quantities obtained as
\begin{eqnarray}
v_c&=&\frac{1}{\kappa}\left[q^2(d-2)(2d-5)\right]^{1/[2(d-3)]},\label{10}
\\
T_c&=&\frac{(d-3)^2}{\pi \kappa v_c(2d-5)},\label{11}
\\
P_c&=&\frac{(d-3)^2}{16\pi \kappa^2v_c^2},\label{12}
\end{eqnarray}%
where $v$ denotes the specific volume and is related to the horizon radius $r_+$ via $r_+=\kappa v$ with $\kappa=\frac{d-2}{4}$. Note that the thermodynamic volume was obtained in Ref. \cite{Cvetic} as
\begin{equation}
V=\frac{\omega_{d-2}r_+^{d-1}}{d-1}. \label{13}
\end{equation}

In the extended phase space, the first law of black hole thermodynamics and the Smarr formula for $d$-dimensional charged AdS black holes reads~\cite{Gunasekaran}
\begin{eqnarray}
dM&=&TdS+\Phi Q+VdP\;, \label{14} \\
M&=&\frac{d-2}{d-3}TS+\Phi Q-\frac{2}{d-3}VP\;. \label{15}
\end{eqnarray}
Here, the mass of the black hole $M$ should be interpreted as the enthalpy $H$ \cite{Kastor}. And Eq. (\ref{14}) can be rewritten as
\begin{equation}
dH=TdS+\Phi dQ+VdP. \label{16}
\end{equation}

\section{Joule-Thomson expansion of $d$-dimensional charged AdS black holes}
\label{Sec3}

Joule-Thomson expansion refers to the expansion of gas from a high pressure section to a low pressure section through a porous plug. The main feature of this process is that the enthalpy is kept constant. To investigate the Joule-Thomson expansion, the Joule-Thomson coefficient serves as an important physical quantity whose sign can be utilized to determine whether heating or cooling will occur. The Joule-Thomson coefficient is defined via the change of the temperature with respect to the pressure as follow
\begin{equation}
\mu=\left(\frac{\partial T}{\partial P}\right)_{H}\;. \label{17}
\end{equation}

\subsection{Two different approaches to derive the Joule-Thomson coefficient}
The first approach utilizes both the first law of black hole thermodynamics and the differentiation of the Smarr formula (Note that similar approach was carried out in the case of Kerr-AdS black holes \cite{Aydiner2}).
Differentiate Eq. (\ref{15}), one can get
\begin{equation}
(d-3)dM=(d-2)TdS+(d-2)SdT+(d-3)\Phi dQ+(d-3)Qd\Phi-2VdP-2PdV\;.  \label{18}
\end{equation}
Utilizing the relations $dM=dH=0$ and $dQ=0$, Eqs. (\ref{16}) and (\ref{18}) can be simplified into
\begin{eqnarray}
0&=&TdS+VdP\;, \label{19} \\
0&=&(d-2)TdS+(d-2)SdT+(d-3)Qd\Phi-2VdP-2PdV\;. \label{20}
\end{eqnarray}
From the two equations above, the Joule-Thomson coefficient can be derived as
\begin{equation}
\mu=\frac{1}{(d-2)S}\left[Vd+2P\left(\frac{\partial V}{\partial P}\right)_{H}-(d-3)Q\left(\frac{\partial \Phi}{\partial P}\right)_{H}\right]\;. \label{21}
\end{equation}
Note that it is convenient to obtain the inversion pressure from this approach by setting $\mu=0$. The inversion pressure $P_i$ reads
\begin{equation}
P_i=\frac{1}{2}\left(\frac{\partial P}{\partial V}\right)_{H}\left[(d-3)Q\left(\frac{\partial \Phi}{\partial P}\right)_{H}-Vd\right]\;. \label{22}
\end{equation}
Utilizing Eqs. (\ref{3}), (\ref{4}), (\ref{8}), (\ref{9}) and (\ref{13}), Eq. (\ref{21}) can be derived as
\begin{equation}
\mu=\frac{4r_+\left[32(5-2d)\pi^2Q^2r_+^{6}+8(d-3)(d-1)M\pi r_+^{d+3}\omega_{d-2}-(d-3)r_+^{2d}\omega_{d-2}^2\right]}{(d-1)\left[-32(d-2)\pi^2Q^2r_+^{6}+8(d-3)(d-1)M\pi r_+^{d+3}\omega_{d-2}-(d-3)(d-2)r_+^{2d}\omega_{d-2}^2\right]}\;. \label{23}
\end{equation}

The second approach only utilizes the first law of black hole thermodynamics. From Eq. (\ref{16}), one can obtain
\begin{equation}
0=T\left(\frac{\partial S}{\partial P}\right)_{H}+V\;. \label{24}
\end{equation}
Note that we have used the conditions $dH=0$ and $dQ=0$. To derive the first term on the righthand side of the above equation, one can consider the entropy $S$
as the function of the Hawking temperature $T$, the pressure $P$ and the charge $Q$. So
\begin{equation}
dS=\left(\frac{\partial S}{\partial P}\right)_{T,Q}dP+\left(\frac{\partial S}{\partial T}\right)_{P,Q}dT+\left(\frac{\partial S}{\partial Q}\right)_{T,P}dQ\;. \label{25}
\end{equation}
Note that the last term can be omitted since $dQ=0$. From Eq. (\ref{24}), one can obtain
\begin{equation}
\left(\frac{\partial S}{\partial P}\right)_{H}=\left(\frac{\partial S}{\partial P}\right)_{T,Q}+\left(\frac{\partial S}{\partial T}\right)_{P,Q}\left(\frac{\partial T}{\partial P}\right)_{H}\;. \label{26}
\end{equation}
Utilizing Eqs. (\ref{24}) and (\ref{26}), one can derive
\begin{equation}
\mu=\frac{1}{T\left(\frac{\partial S}{\partial T}\right)_{P,Q}}\left[T\left(\frac{\partial V}{\partial T}\right)_{P,Q}-V\right]=\frac{1}{C_P}\left[T\left(\frac{\partial V}{\partial T}\right)_{P,Q}-V\right]\;. \label{27}
\end{equation}
In the above derivation, we have used the definition of $C_P$ and the Maxwell relation. It is convenient to obtain the inversion temperature from this approach by setting $\mu=0$. The inversion temperature $T_i$ reads
\begin{equation}
T_i=V\left(\frac{\partial T}{\partial V}\right)_{P,Q}\;. \label{28}
\end{equation}

From Eqs. (\ref{7}), (\ref{8}) and (\ref{13}), Eq. (\ref{27}) can be obtained as
\begin{equation}
\mu=\frac{4r_+\left\{(d-3)[dr_+^{2d}-3(d-2)q^2r_+^6]+16P\pi r_+^{2d+2}\right\}}{(d-1)\left\{(d-2)(d-3)[r_+^{2d}-q^2r_+^6]+16P\pi r_+^{2d+2}\right\}}\;. \label{29}
\end{equation}

Utilizing Eqs. (\ref{3}), (\ref{4}) and (\ref{5}), one can prove that Eq. (\ref{23}) is exactly in accord with Eq. (\ref{29}). Then we can conclude that both approaches are consistent with each other and one may choose either approach to investigate the Joule-Thomson expansion of charged AdS black holes.

\subsection{Effects of dimensionality on the Joule-Thomson coefficient, inversion curves and isenthalpic curves}

\begin{figure}[H]
\centerline{\subfigure[]{\label{1a}
\includegraphics[width=8cm,height=6cm]{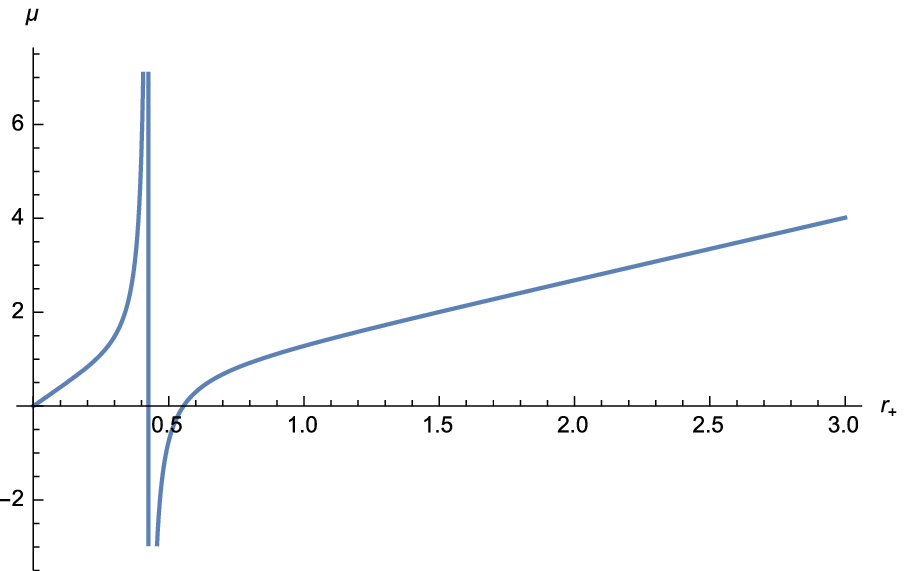}}
\subfigure[]{\label{1b}
\includegraphics[width=8cm,height=6cm]{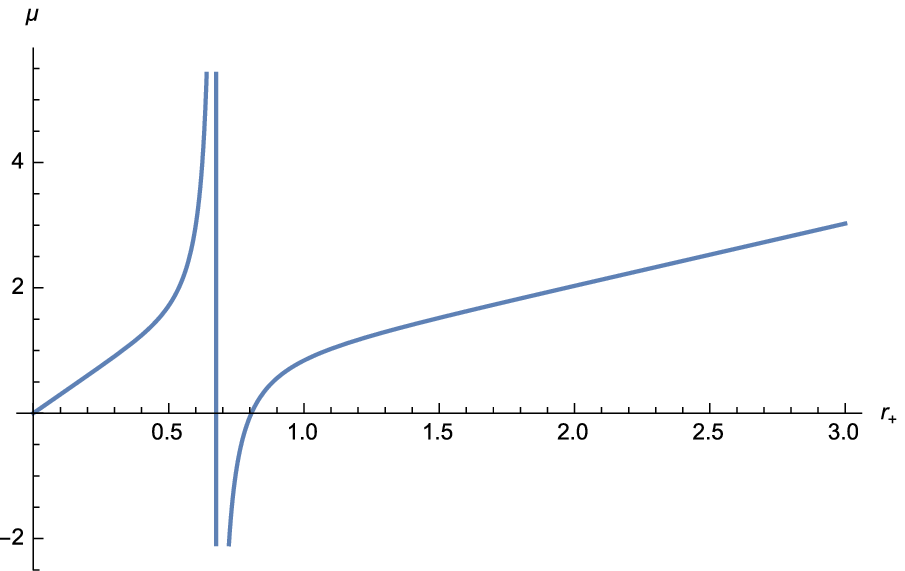}}}
\centerline{\subfigure[]{\label{1c}
\includegraphics[width=8cm,height=6cm]{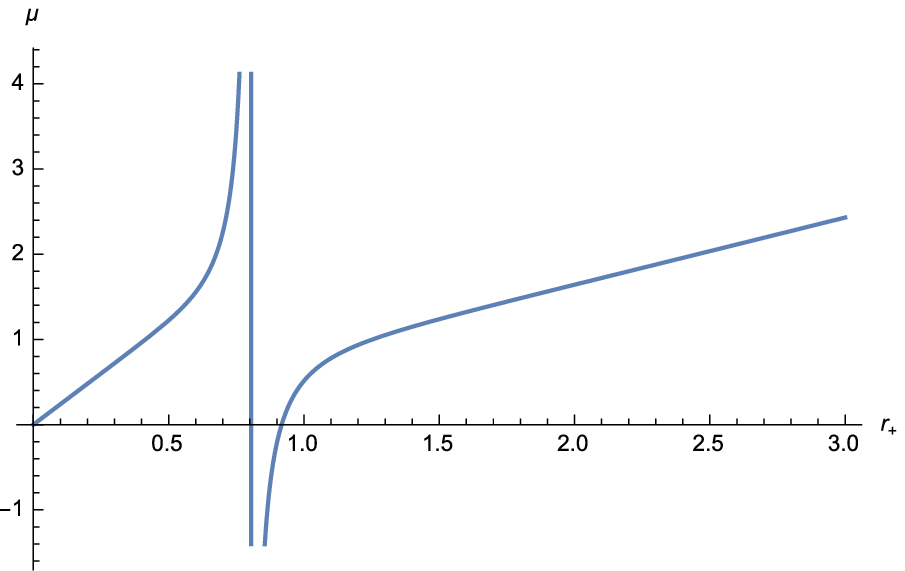}}
\subfigure[]{\label{1d}
\includegraphics[width=8cm,height=6cm]{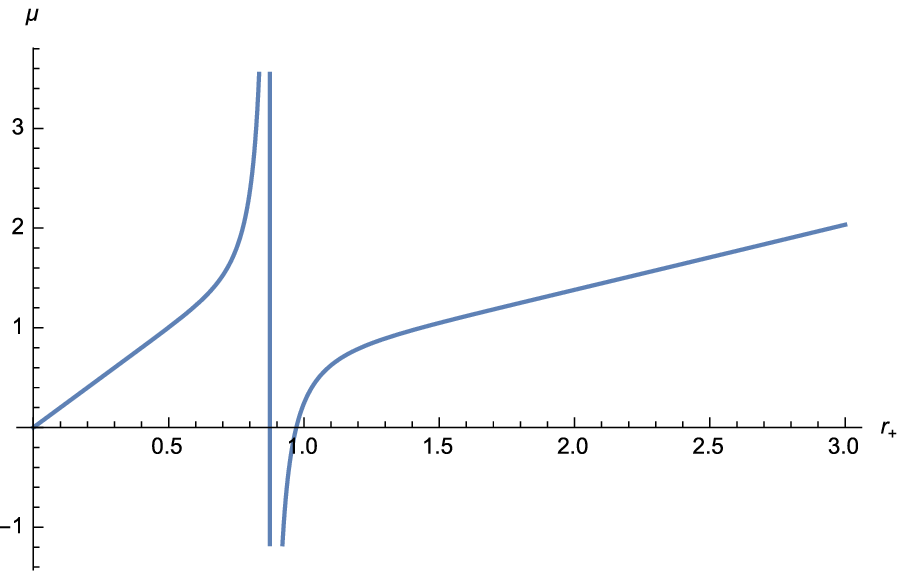}}}
\centerline{\subfigure[]{\label{1e}
\includegraphics[width=8cm,height=6cm]{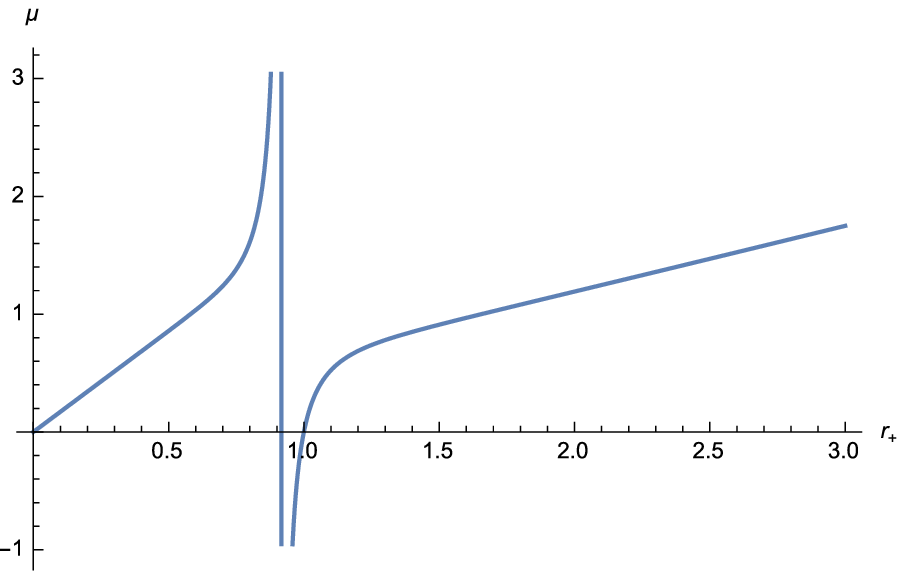}}
\subfigure[]{\label{1f}
\includegraphics[width=8cm,height=6cm]{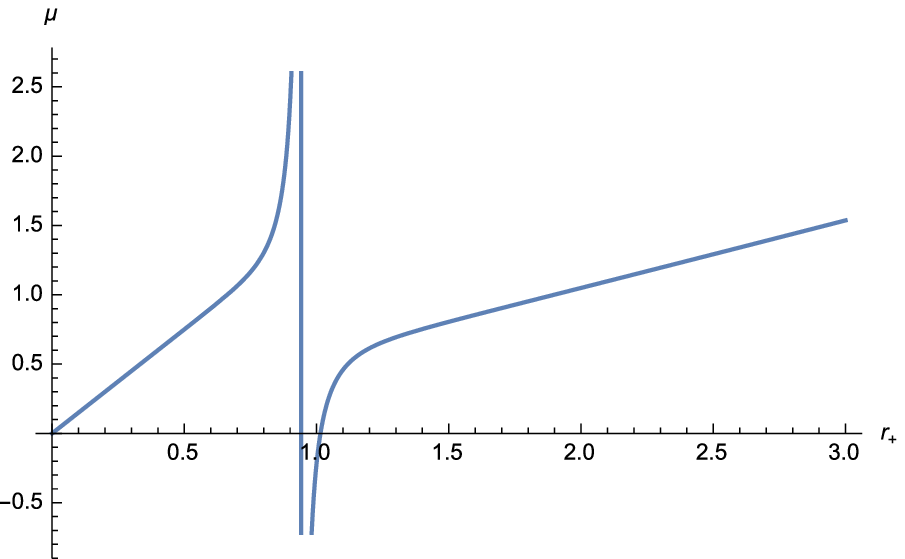}}}
 \caption{Joule-Thomson coefficient $\mu$ for $P=1, q=1$ (a) $d=4$ (b) $d=5$ (c) $d=6$ (d) $d=7$ (e) $d=8$ (f) $d=9$}
\label{fg1}
\end{figure}

The effect of dimensionality on the behavior of $\mu$ can be observed from Fig.\ref{fg1}, which we have utilized Eq. (\ref{29}) to plot. There exist both a divergent point and a zero point in each subgraph. And the corresponding horizon radius increases with the dimension $d$.

\begin{figure}[H]
\centerline{\subfigure[]{\label{2a}
\includegraphics[width=8cm,height=6cm]{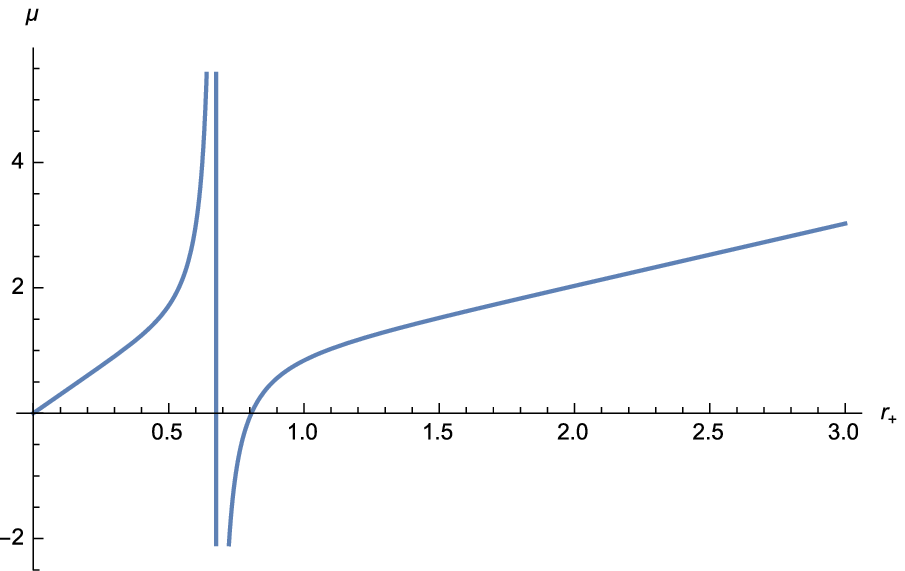}}
\subfigure[]{\label{2b}
\includegraphics[width=8cm,height=6cm]{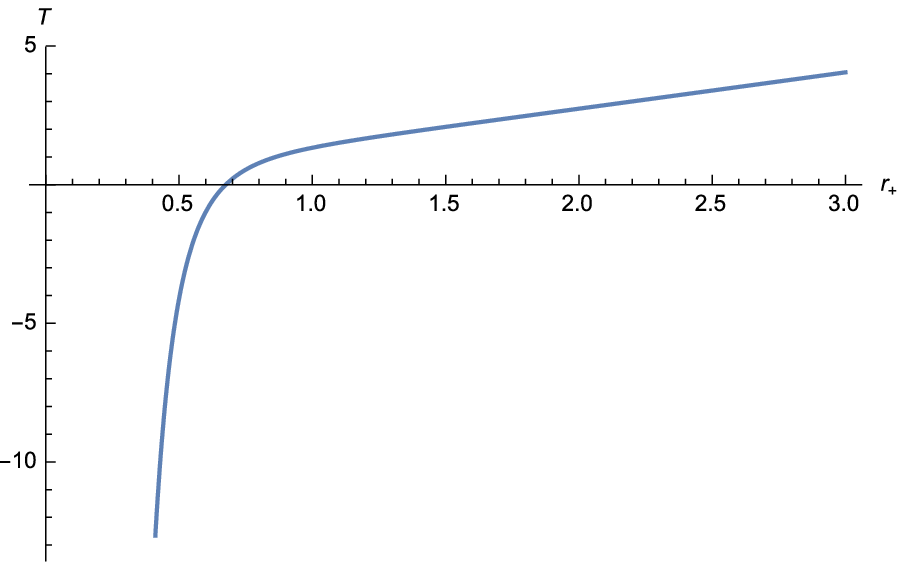}}}
 \caption{ (a)Joule-Thomson coefficient $\mu$ for $P=1, q=1$, $d=5$ (b)Hawking temperature $T$ for $P=1, q=1$, $d=5$ }
\label{fg2}
\end{figure}

The divergent point here reveals the information of Hawking temperature and corresponds to the extremal black hole. To further understand this issue, we present the graph of the Joule-Thomson coefficient $\mu$ and the Hawking temperature $T$ for the case $d=5$ in Fig.\ref{fg2}. One can see clearly that the divergent point of Joule-Thomson coefficient coincides with the zero point of Hawking temperature.

The zero point of Joule-Thomson coefficient is of great physical significance in the study of Joule-Thomson Expansion. It is the inversion point that discriminate the cooling process from heating process. According to the definition of Joule-Thomson coefficient (Eq. (\ref{17})), $\mu>0$ corresponds to the cooling process while $\mu<0$ corresponds to the heating process. At the first glance of Eq. (\ref{27}), one may suspect that Joule-Thomson coefficient equals to zero when the specific heat $C_P$ diverge. This will lead to interesting findings since $C_P$ have two divergent points when the electric charge is smaller than the critical value. However, the reality is not so good as we imagine. Eq. (\ref{27}) can be further simplified into
\begin{equation}
\mu=\left(\frac{\partial V}{\partial S}\right)_{P,Q}-\frac{V}{T\left(\frac{\partial S}{\partial T}\right)_{P,Q}}=\left(\frac{\partial V}{\partial S}\right)_{P,Q}-\frac{V}{C_P}\;. \label{30}
\end{equation}
It can be witnessed that the Joule-Thomson coefficient does not equal to zero even if $C_P$ diverge due to the nonzero contribution of the first term on the right hand side of the above equation. Eq. (\ref{30}) also explains quite well why the Joule-Thomson coefficient diverges when the Hawking temperature approaches zero.

\begin{figure}[H]
\centerline{\subfigure[]{\label{3a}
\includegraphics[width=8cm,height=6cm]{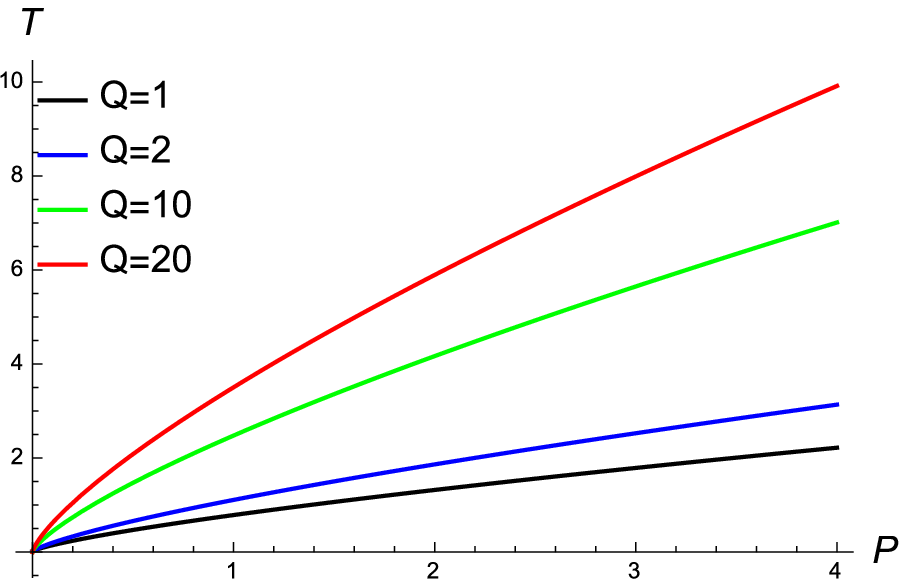}}
\subfigure[]{\label{3b}
\includegraphics[width=8cm,height=6cm]{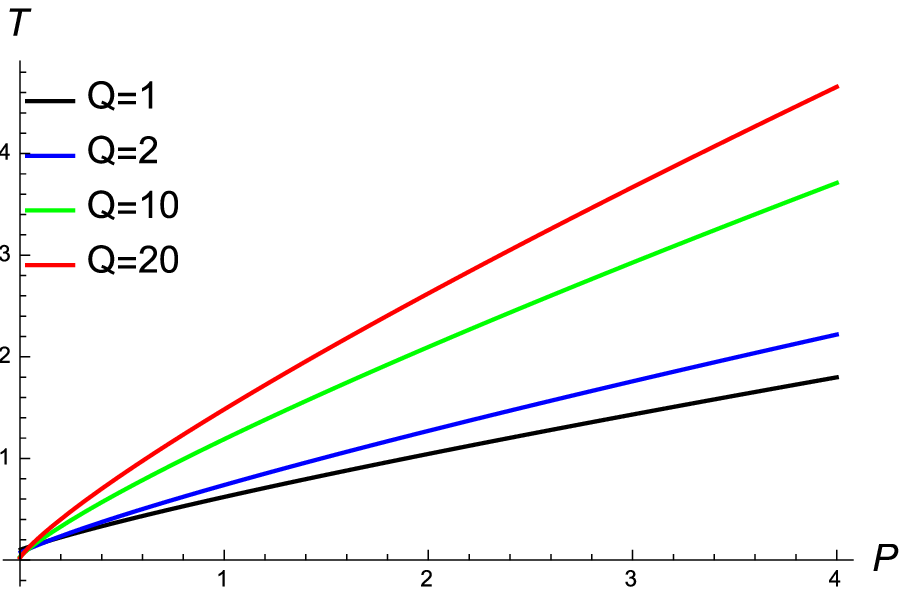}}}
\centerline{\subfigure[]{\label{3c}
\includegraphics[width=8cm,height=6cm]{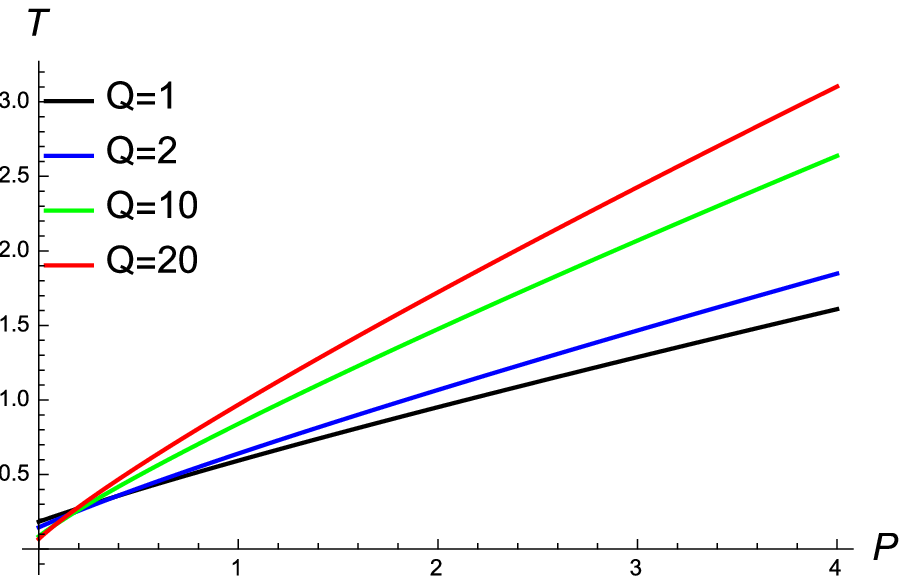}}
\subfigure[]{\label{3d}
\includegraphics[width=8cm,height=6cm]{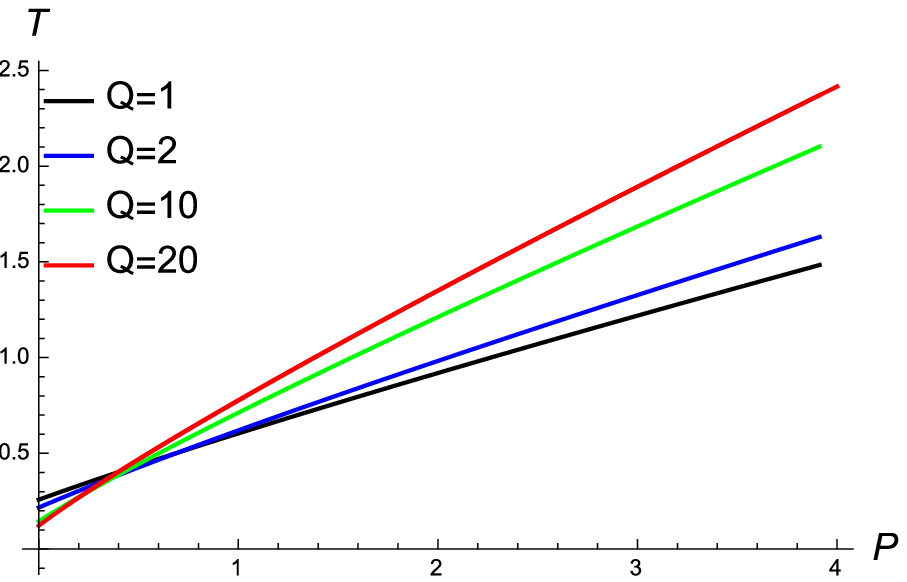}}}
\centerline{\subfigure[]{\label{3e}
\includegraphics[width=8cm,height=6cm]{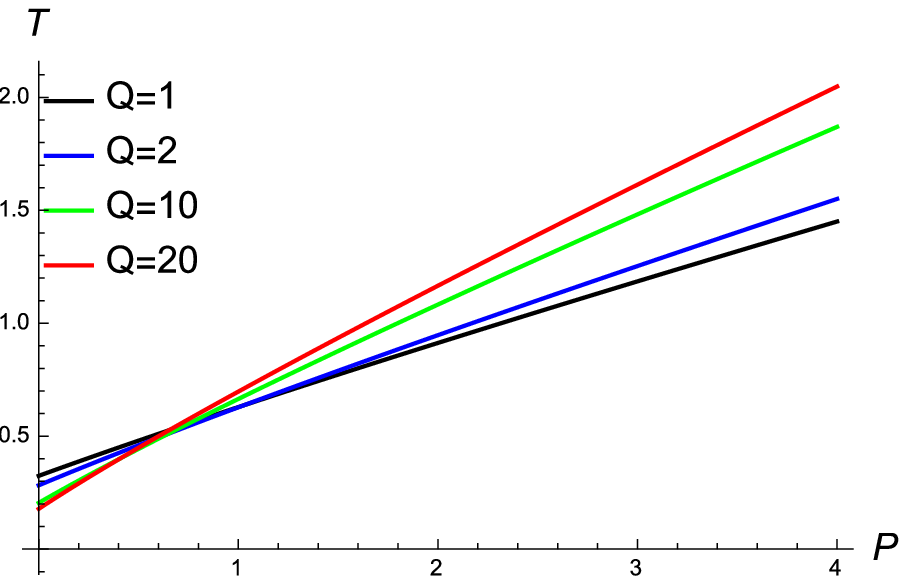}}
\subfigure[]{\label{3f}
\includegraphics[width=8cm,height=6cm]{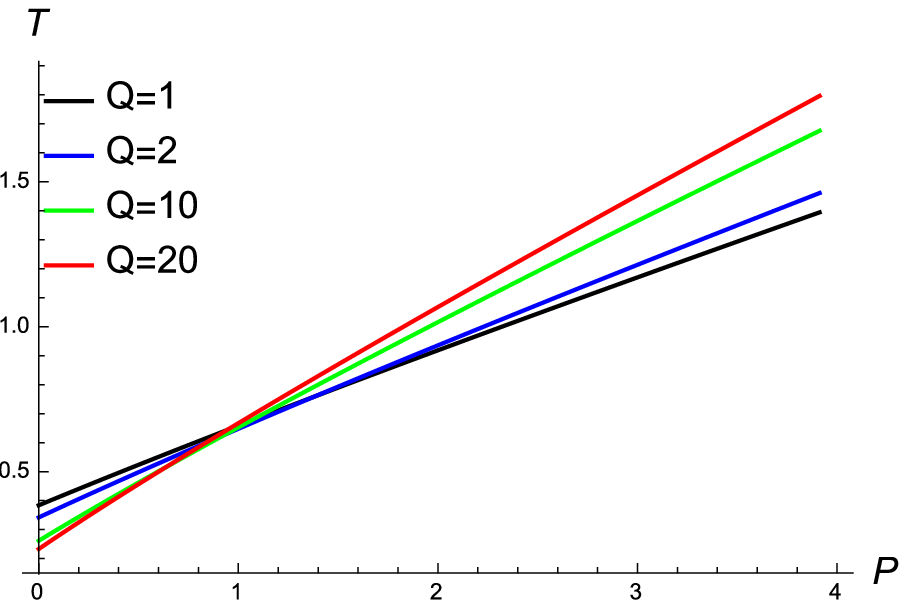}}}
 \caption{The inversion curve for (a) $d=4$ (b) $d=5$ (c) $d=6$ (d) $d=7$ (e) $d=8$ (f) $d=9$}
\label{fg3}
\end{figure}

From Eq. (\ref{29}), one can conclude that the inversion pressure $P_i$ and the corresponding $r_{+i}$ satisfy the following relation
 \begin{equation}
(d-3)[dr_{+i}^{2d}-3(d-2)q^2r_{+i}^6]+16P_i\pi r_{+i}^{2d+2}=0\;. \label{31}
\end{equation}
The explicit expressions of the solutions for $d>4$ are so lengthy that we will not list them here. Instead, we will depict the relation between the inversion temperature $T_i$ and the inversion pressure $P_i$ with the help of Eq. (\ref{7}). As can be seen from Fig.\ref{fg3}, the inversion curves for $d>4$ behave similarly as that of the case $d=4$ \cite{Aydiner1}. There exists only one branch in the inversion curve. The inversion temperature increases monotonically with the inversion pressure, although the slope of the inversion curves decreases. Fine structures can be observed in the cases $d>4$. For low pressure, the inversion temperature decreases with the electric charge $Q$ while it increases with $Q$ for high pressure.


\begin{figure}[H]
\centerline{\subfigure[]{\label{4a}
\includegraphics[width=8cm,height=6cm]{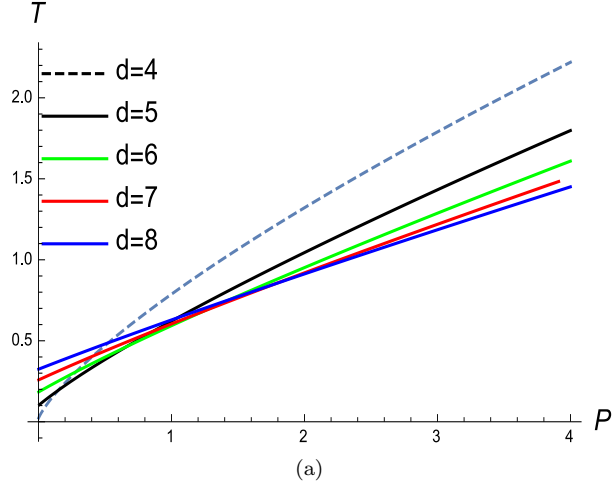}}}
 \caption{Effect of the dimensionality on the inversion curve for $Q=1$}
\label{fg4}
\end{figure}

To compare the effect of the dimensionality more intuitively, we show the inversion curves of $d$-dimensional charged AdS black holes for $Q=1$ in Fig.\ref{fg4}. At low pressure, the inversion temperature increases with the dimensionality $d$ while at high pressure it decreases with $d$.

It is of interest to probe the ratio between minimum inversion temperature $T_{min}$ and the critical temperature $T_c$. It was reported that this ratio turns out to be $1/2$ for four-dimensional charged AdS black holes \cite{Aydiner1} and Kerr-AdS black holes \cite{Aydiner2}. Note that $T_{min}$ can be obtained by demanding $P_i=0$.
 \begin{equation}
T_{min}=\frac{2^{\frac{1-2d}{2(d-3)}}3^{\frac{5-2d}{2(d-3)}}\pi^{\frac{2-d}{d-3}}(d-3)^2}{d-2}\left[\frac{Q^2}{\omega_{d-2}^2d(d-3)}\right]^{\frac{1}{6-2d}}\;. \label{32}
\end{equation}
When $d=4$, Eq. (\ref{32}) reduces to be
 \begin{equation}
T_{min}|_{d=4}=\frac{1}{6\sqrt{6}\pi Q}\;,\label{33}
\end{equation}
recovering the result of Ref. \cite{Aydiner1}.
Utilizing Eqs. (\ref{5}), (\ref{10}), (\ref{11}) and (\ref{32}), the ratio of $T_{min}/T_c$ can be obtained as
 \begin{equation}
\frac{T_{min}}{T_c}=\frac{3^{\frac{5-2d}{2(d-3)}}(2d-5)[d(2d-5)]^{\frac{1}{2(d-3)}}}{2(d-2)}\;,\label{33}
\end{equation}
Based on Eq. (\ref{33}), we list the ratio $T_{min}/T_c$ for various dimensions in Table \ref{tb1}. For $d=4$, it recovers the result of the former literature \cite{Aydiner1}. However, for $d>4$, the ratio is not always equal to $1/2$ and it decreases with the dimensionality $d$.
\begin{table}[!h]
\tabcolsep 0pt
\caption{The ratio $T_{min}/T_c$ for various dimensions}
\vspace*{-12pt}
\begin{center}
\def\temptablewidth{0.5\textwidth}
{\rule{\temptablewidth}{1pt}}
\begin{tabular*}{\temptablewidth}{@{\extracolsep{\fill}}ccccccc}
$d$ & 4 & 5 &6 &7 &8 &9 \\   \hline
    $T_{min}/T_c$ \; & 1/2 \;&0.471957 \;&      0.452802\;& 0.438933\;&0.428377\; &0.420032  \\
       \end{tabular*}
       {\rule{\temptablewidth}{1pt}}
       \end{center}
       \label{tb1}
       \end{table}

\begin{figure}[H]
\centerline{\subfigure[]{\label{5a}
\includegraphics[width=8cm,height=6cm]{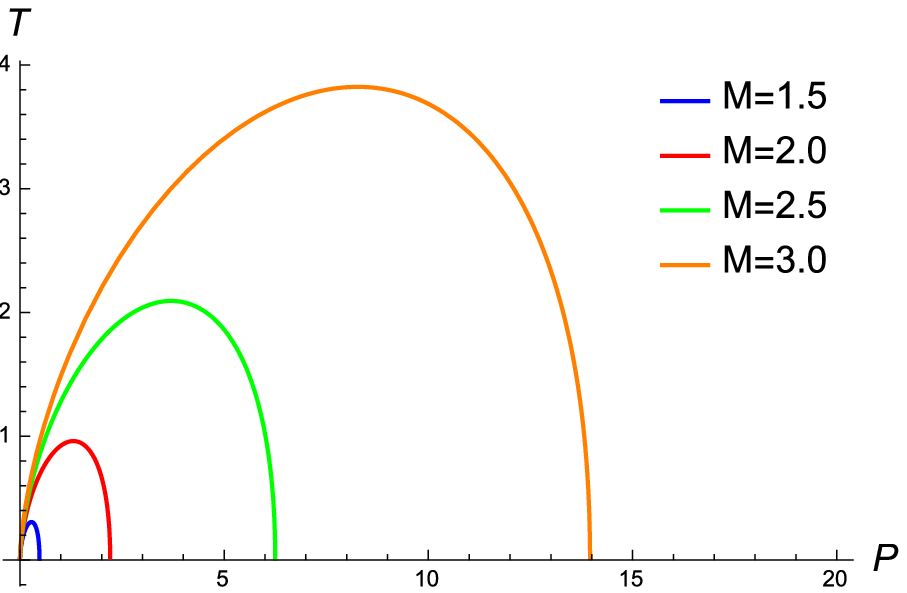}}
\subfigure[]{\label{5b}
\includegraphics[width=8cm,height=6cm]{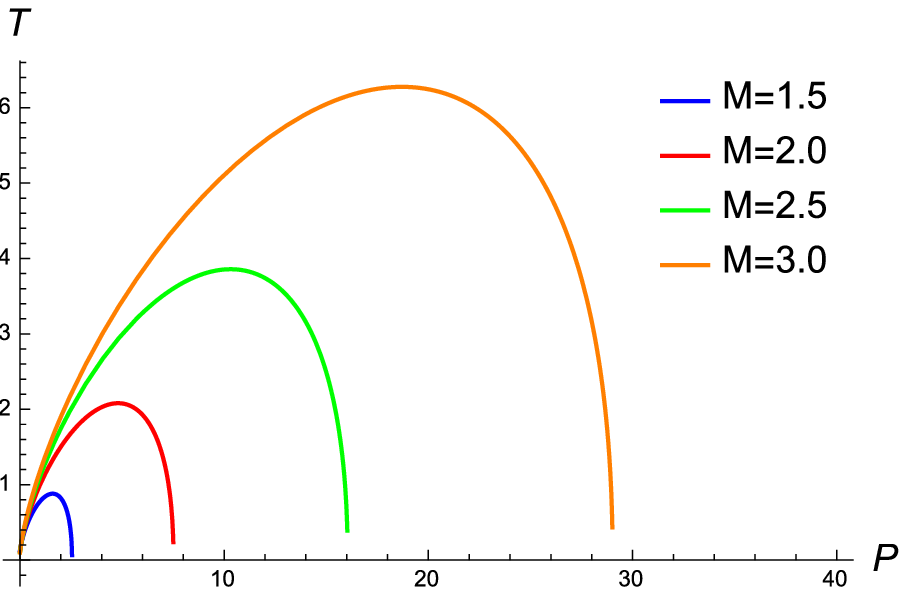}}}
\centerline{\subfigure[]{\label{5c}
\includegraphics[width=8cm,height=6cm]{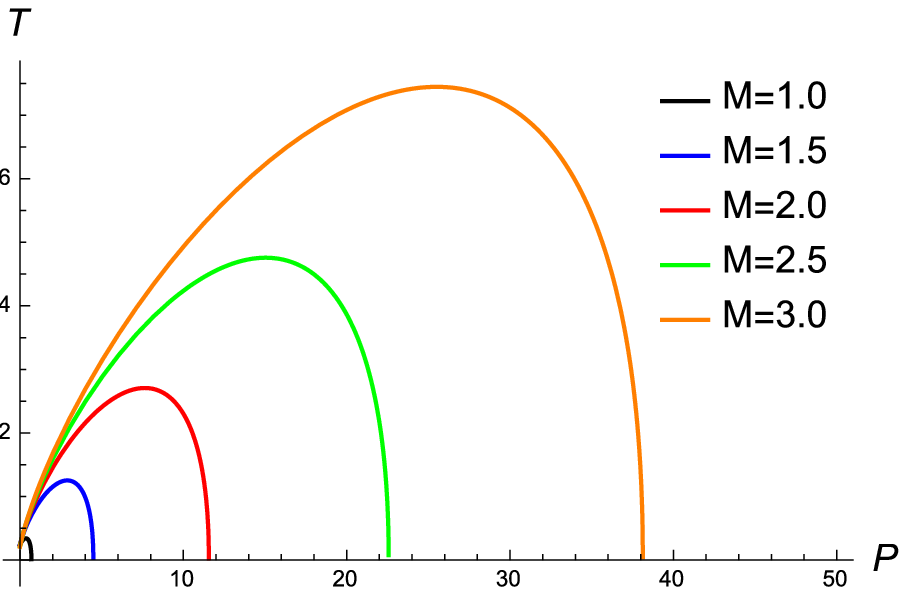}}
\subfigure[]{\label{5d}
\includegraphics[width=8cm,height=6cm]{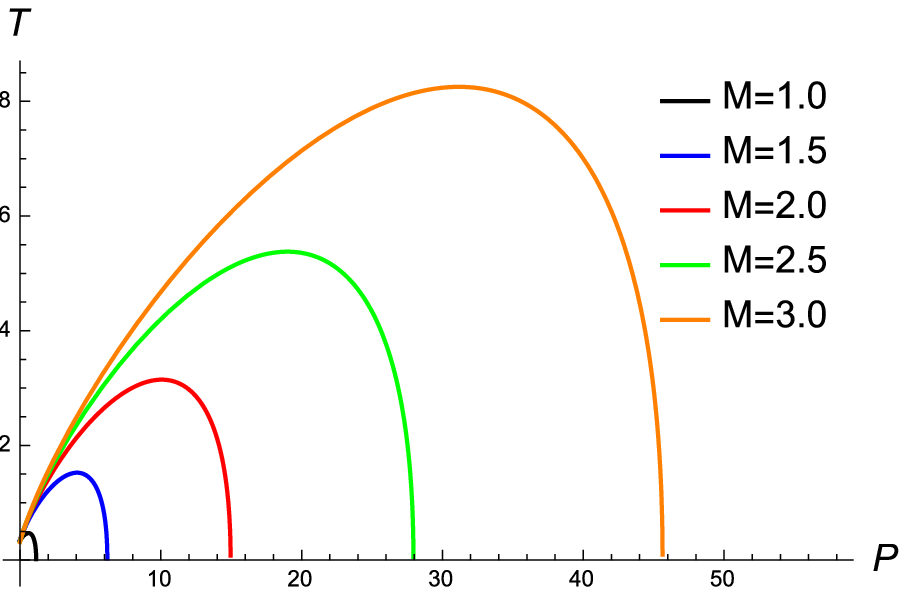}}}
\centerline{\subfigure[]{\label{5e}
\includegraphics[width=8cm,height=6cm]{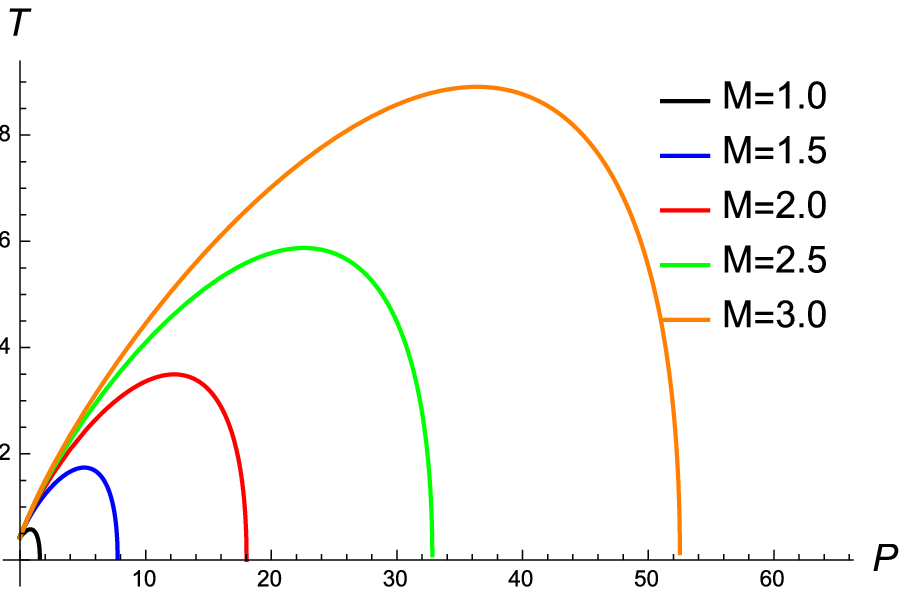}}
\subfigure[]{\label{5f}
\includegraphics[width=8cm,height=6cm]{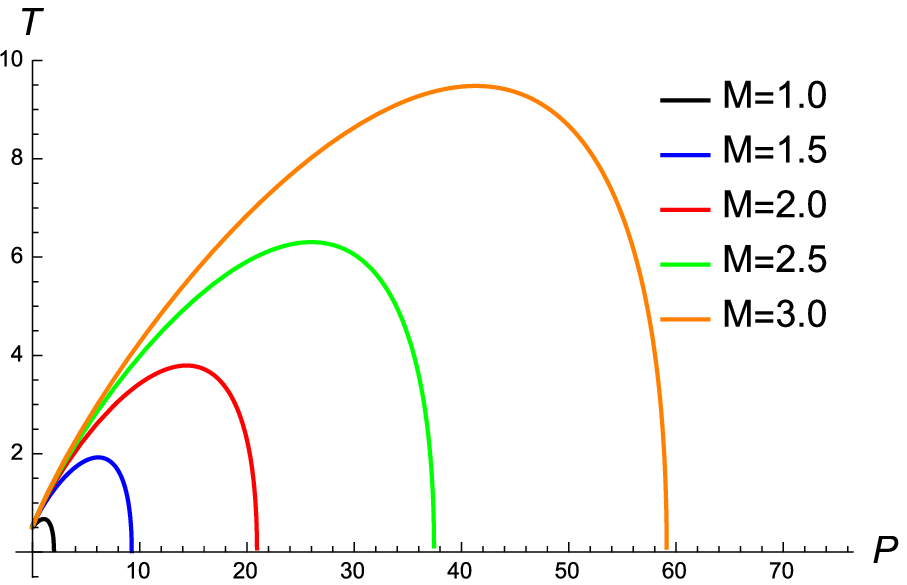}}}
 \caption{The isenthalpic curve for (a) $d=4$ (b) $d=5$ (c) $d=6$ (d) $d=7$ (e) $d=8$ (f) $d=9$ with $Q$ chosen as $Q=1$}
\label{fg5}
\end{figure}

Since the Joule-Thomson expansion is a process during which the enthalpy is kept constant, it would also be interesting to study the isenthalpic curves of $d$-dimensional charged AdS black holes. In the extended phase space, the mass should be interpreted as enthalpy. So isenthalpic curves can be depicted if one fix the mass of the black hole. Utilizing Eqs. (\ref{3}), (\ref{4}) and (\ref{7}), we plot the isenthalpic curves for various values of mass.
As can be seen from Fig.\ref{fg5}, the isenthalpic curves of $d>4$ look similar to the curves of $d=4$ \cite{Aydiner1} except that the curves tend to expand rightward (toward higher pressure) when the dimensionality $d$ increases.

\section{Conclusions}
\label{Sec4}
In this paper, we discuss in detail the effects of the dimensionality on the Joule-Thomson expansion by considering the case of $d$-dimensional charged AdS black holes. The Joule-Thomson expansion describes the expansion of gas from a high pressure section to a low pressure section through a porous plug.

Firstly, we investigate the famous Joule-Thomson coefficient $\mu$ because it serves as an important physical quantity whose sign can be utilized to determine whether heating or cooling will occur. We apply two different approaches to derive the Joule-Thomson coefficient. The first approach utilizes both the first law of black hole thermodynamics and the differentiation of the Smarr formula while the second approach only utilizes the first law of black hole thermodynamics. For both approaches, we derive the explicit expression of $\mu$, from which we show that both approaches are consistent with each other and one may choose either approach to investigate the Joule-Thomson expansion of charged AdS black holes. We also plot the behavior of $\mu$ for various dimensions. There exist both a divergent point and a zero point with the corresponding horizon radius increasing with the dimensionality $d$. The divergent point of Joule-Thomson coefficient coincides with the zero point of Hawking temperature and hence reveals the information of Hawking temperature. The zero point of Joule-Thomson coefficient is the inversion point that discriminate the cooling process ($\mu>0$)from heating process ($\mu<0$). By simplifying the expression of $\mu$, we show that the Joule-Thomson coefficient does not equal to zero even if $C_P$ diverge.

Secondly, we study the inversion curves and relevant physical quantities. It is shown graphically that the inversion curves for $d>4$ behave similarly as that of the case $d=4$. There exists only one branch in the inversion curve. The inversion temperature increases monotonically with the inversion pressure, although the slope of the inversion curves decreases. Moreover, we disclose fine structures in the cases $d>4$. For low pressure, the inversion temperature decreases with the electric charge $Q$ while it increases with $Q$ for high pressure. The effect of the dimensionality is quite the reverse. At low pressure, the inversion temperature increases with the dimensionality $d$ while at high pressure it decreases with $d$. The ratio between minimum inversion temperature $T_{min}$ and the critical temperature $T_c$ is discussed with its explicit expression obtained for $d>4$. For $d=4$, it recovers the result of the former literature \cite{Aydiner1}. However, for $d>4$, it is shown that the ratio is not always equal to $1/2$ but decreases with the dimensionality $d$.

Last but not the least, we focus on the isenthalpic curves for various dimensions since the main feature of the Joule-Thomson expansion is that the enthalpy is kept constant. In the extended phase space, the mass should be interpreted as enthalpy. So we depict isenthalpic curves by fixng the mass of the black hole. It is shown that the isenthalpic curves of $d>4$ look similar to the curves of $d=4$ except that the curves tend to expand rightward (toward higher pressure) when the dimensionality $d$ increases.

 \section*{Acknowledgements}

 This research is supported by National Natural Science Foundation of China (Grant No.11605082). The authors are also in part supported by National Natural Science Foundation of China (Grant No.11747017), Natural Science Foundation of Guangdong Province, China (Grant Nos.2016A030310363, 2016A030307051, 2015A030313789) and Department of Education of Guangdong Province, China.

\end{document}